\documentclass[prl,twocolumn,showpacs,letterpaper,showpacs,superscriptaddress]{revtex4-1}
\usepackage{graphicx,amsmath,amssymb,amsfonts,latexsym,color,dcolumn,bm,subfigure}

\begin{document}

\title{Limits on the accuracy of isoelectronic gravity measurements at short separation due to patch potentials}

\author{R. O. Behunin}
\affiliation{Center for Nonlinear Studies, Los Alamos National Laboratory, Los Alamos, New Mexico 87545, USA}
\affiliation{Theoretical Division, MS B213, Los Alamos National Laboratory, Los Alamos, New Mexico 87545, USA}
\author{D. A. R. Dalvit}
\affiliation{Theoretical Division, MS B213, Los Alamos National Laboratory, Los Alamos, New Mexico 87545, USA}
\author{R. S. Decca}
\affiliation{Department of Physics, Indiana University-Purdue University Indianapolis, Indianapolis, Indiana 46202, USA}
\author{C. C. Speake}
\affiliation{Gravitation Group, School of Physics and Astronomy,  University of Birmingham, Birmingham, United Kingdom}

\date{ \today}

\begin{abstract}
In force sensing experiments intended to measure non-Newtonian gravitational signals electrostatic patch potentials can give rise to spurious forces, torques, and noise. Undesired patch-induced interactions can lead to systematic effects which limit accuracy, and noise can place lower limits on precision.  In this paper we develop the theory for electrostatic patch effects on isoelectronic experiments, where their mean effect is nullified by design.  We derive analytical expressions for the patch force and torque power spectrum to estimate the limitations introduced by patch-induced signals.
\end{abstract}


\maketitle

The presence of patch potentials on metallic surfaces has important implications in various experiments, including 
measurements of gravity on elementary particles \cite{Fairbank},
heating in ion traps  \cite{Deslauriers06,Hite12}, ionization of Rydberg atoms \cite{Pu10}, precision tests of general relativity in space \cite{Pollack08,Everitt11},  measurements of the Casimir force \cite{Decca05b,Sushkov11a,Mohideen12,Garcia12}, and searches for
hypothetical forces \cite{Fischbach98,Long,Adelberger03,Speake04,Decca05,Kapner05,Kapner07,Kapitulnik,Masuda,Sushkov11b,Antoniadis12}. Patches lead to interactions that limit the accuracy and precision of such experiments. To date, the theoretical modeling has focussed on quantifying the mean effect of patches on measured signals \cite{Speake03,Kim10,Carter11,Behunin12,Behunin12b}, but an in-depth analysis of the patch force/torque power spectrum is lacking.  
Here, we develop such a theory and apply it to the analysis of two ongoing experiments that aim to measure new hypothetical forces in which the mean patch effect is cancelled out by construction, but where variations of the surface potential still detrimentally affect the ability to constrain new physics.  

These two experiments, one being performed at the University of Birmingham \cite{Clive-new-exp} and another at IUPUI \cite{Ricardo-new-exp},  use the isoelectronic technique. In this approach a flat material sample with inhomogeneous mass density is covered with a sufficiently thick metal coating to electrically screen the underlying materials. This has two important consequences: 
Both (i) the Casimir interaction and (ii) the mean patch force  between the coated sample and a force probe are independent of the location of the latter at fixed separation distance. Force signals that depend on lateral displacements can arise not only from the sought after mass-dependent interactions between the probe and the underlying inhomogeneous mass density, but from spatially dependent patch force fluctuations as well. 
Hence, although the isoelectronic technique eliminates Casimir and mean patch forces, it is still possible 
that patch effects contribute to the measured signal.  In order to quantify the magnitude of patch systematics, we compute the electrostatic torque and force power spectrum for the above two experiments and show that, for reasonable estimations of parameters describing the sample's surface potentials, patch effects can be an important limitation to accuracy. 


{\it The Birmingham Experiment and Patch Torque:}
This experiment searches for violations of the inverse square law of gravity by measuring the torque acting between two parallel disks \cite{Clive-new-exp}. As shown schematically in Fig. 1a, two identical disks (radius $R_0=40$ mm, thickness 10 $\mu$m)
are fabricated with a pattern of 2048 pairs of gold/copper spokes, and over-laid with a gold film  (thickness 1 $\mu$m, not shown in Fig.1a) 
for isoelectronic measurements. The 2048 mass pairs are grouped into 16 sectors of alternating phase to provide a further modulation of the signal. In the schematic in Figure 1a, for clarity, only 36 mass pairs are shown in 4 sectors of alternating phase.
The lower disk, the test mass, is attached to a torsion pendulum with a superconducting suspension and the upper disk, the source mass, is attached to a micro-positioner. The spacing between the gold surfaces is nominally $D=13 \; \mu$m. 
The experiment has been designed to achieve a sensitivity 
of $10^{-17}$ N m after an integration period of 1 day. The experimental procedure comprises the measurement of the torque as a function
of the rotation angle, which should be angle periodic with a frequency determined by the mass density modulation.

To  compute the patch torque power spectrum we first obtain an expression for the patch torque as a function
of angular displacement $\phi$ between the disks. Given that $D \ll R_0$, we approximate the disks by infinite parallel planes and neglect
finite size effects. The electrostatic interaction energy as a function of $\phi$  for given fixed patch potentials $V_a({\bf x})$ on the disks ($a=1,2$) located at $z_a=0,D$, is
\begin{eqnarray}
\label{Energy}
\mathcal{E}(\phi)  &=&  \int d^2 x \int d^2 x'   \bigg[ V_1({\bf x}) \Delta( \tilde{\bf x} - {\bf x'},D) V_2({\bf x}') 
\nonumber \\
&& + \bigg( V_1({\bf x})  \Sigma({\bf x} - {\bf x'},D) V_1({\bf x}')  +(1 \rightarrow 2)  \bigg)    \bigg] .
\end{eqnarray}
Here $ \Delta({\bf x},D) \equiv - \varepsilon_0 \int \frac{d^2 k}{(2\pi)^2} \ \frac{k}{\sinh k D}e^{ i {\bf k} \cdot {\bf x} }$,
$\Sigma({\bf x},D) \equiv \frac{\varepsilon_0}{2} \int \frac{d^2 k}{(2\pi)^2} \ \frac{k e^{- k D}}{\sinh k D}e^{ i {\bf k} \cdot {\bf x} }$, $\tilde{\bf x} = R^{-1}_\phi \cdot {\bf x}$, ${\bf x}$ is a vector in the $xy-$ plane, $k=|{\bf k}|$, and $\varepsilon_0$ is the permittivity of free space. The matrix $R^{-1}_\phi $ rotates vectors lying in the $xy$-plane (see supplementary information).
The torque is $\tau(\phi) = \partial \mathcal{E}(\phi)/\partial \phi$. Notice that terms containing $\Sigma$ do not contribute to the torque because they are independent of the relative angle.

Without a precise knowledge of the surface potentials $V_a({\bf x})$, we proceed by making some statistical arguments. 
For this purpose we adopt the ``quasi-local" model of random surface potentials \cite{Behunin12}. A given micro-realization
of surface voltages in this model corresponds to a random tessellation of the surface with random and statistically independent voltage assignments to each patch domain. In the following we assume that the patch voltages fluctuate around zero, as a constant background potential does not contribute to the signal in the isoelectronic approach.
Provided many elementary patch areas are contained on the plates we can 
replace the product of surface voltages $V_a({\bf x}) V_b({\bf x}')$  by the two-point voltage correlation function, 
$\langle V_a({\bf x}) V_b({\bf x}') \rangle \equiv  \delta_{ab} C_{aa}({\bf x},{\bf x}')$ (ergodic hypothesis). The angled brackets denote ensemble averaging over all realizations of voltages and geometrical patch layouts. 
Consequently, $\langle V_2({\bf x}) V_1({\bf x}') \rangle =0$ and $\langle \tau(\phi) \rangle = 0$.
Because the tesselation of the surface is determined randomly the voltage correlation exhibits translational and rotational invariance, i.e. $C_{aa}({\bf x},{\bf x}') = C_{aa}(r)$, with $r=|{\bf x}-{\bf x}'|$. The explicit form of the quasi-local correlation function is given in Eq.(14) of \cite{Behunin12}.
For the special case when all patches have the same typical size $\ell$, 
and both disks have the same voltage correlation properties, 
this is given by
$C_{ab}(r) =\delta_{ab} \frac{2 V_{\rm rms}^2}{\pi} [ 
\cos^{-1} \left(\frac{r}{\ell}\right) - 
\frac{r}{\ell} (1- (\frac{r}{\ell})^2)^{1/2}]$,
where $V_{\rm rms}$ is the root-mean-square of the patch voltage.
Unfortunately this correlator is inexpedient for the purposes of computation.
Instead, we approximate it by a Gaussian 
\begin{equation}
C_{aa}({\bf x},{\bf x}') \approx V^2_{\rm rms} \exp(- 4|{\bf x} - {\bf x}'|^2/  \ell^2), 
\label{corr}
\end{equation}
which captures the salient features of the quasi-local correlator and allows an almost analytical treatment (see inset of Fig.1b).

\begin{figure}[t]
\begin{center}
\includegraphics[width=3in]{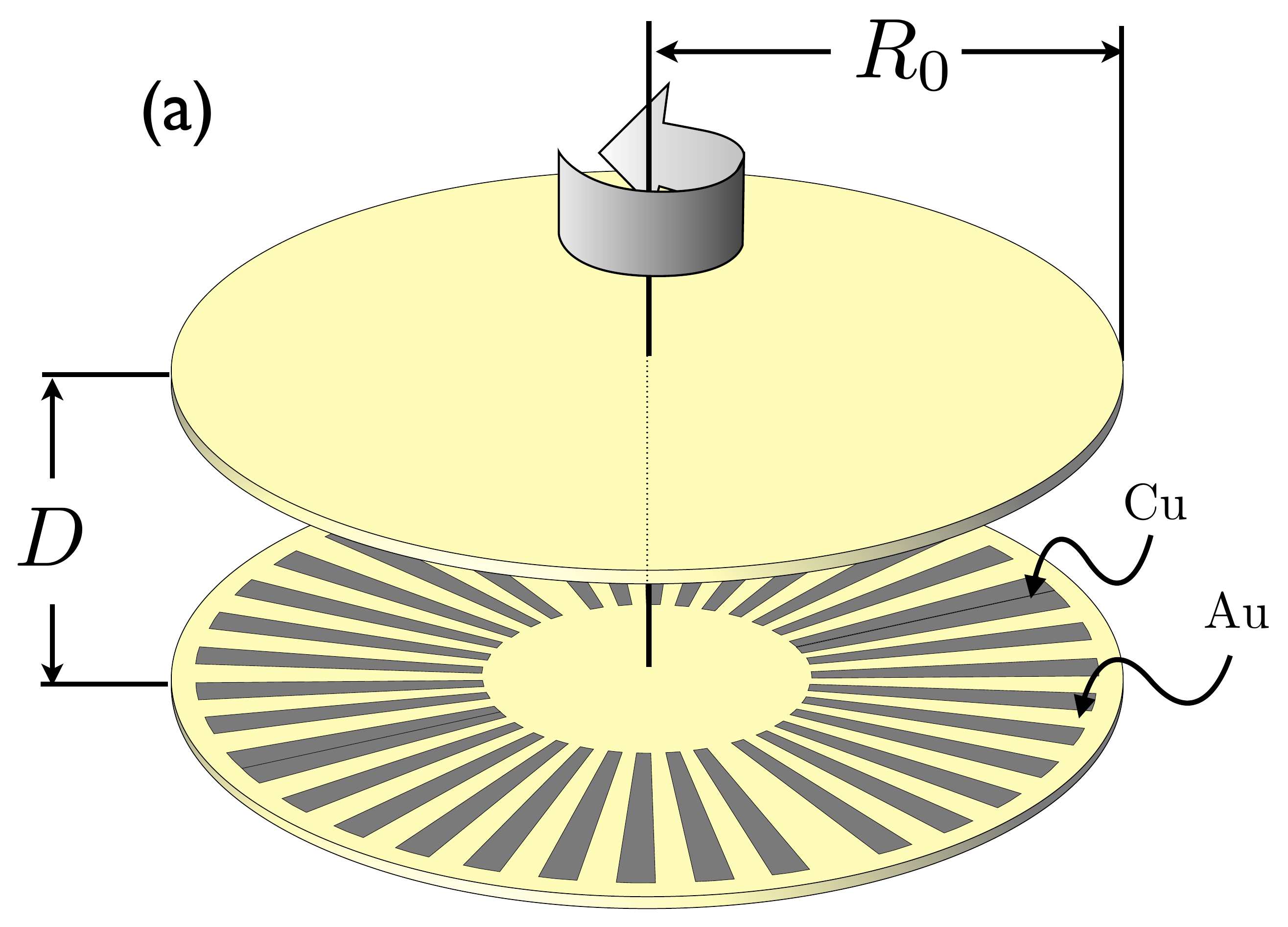}
\includegraphics[width=3.0in]{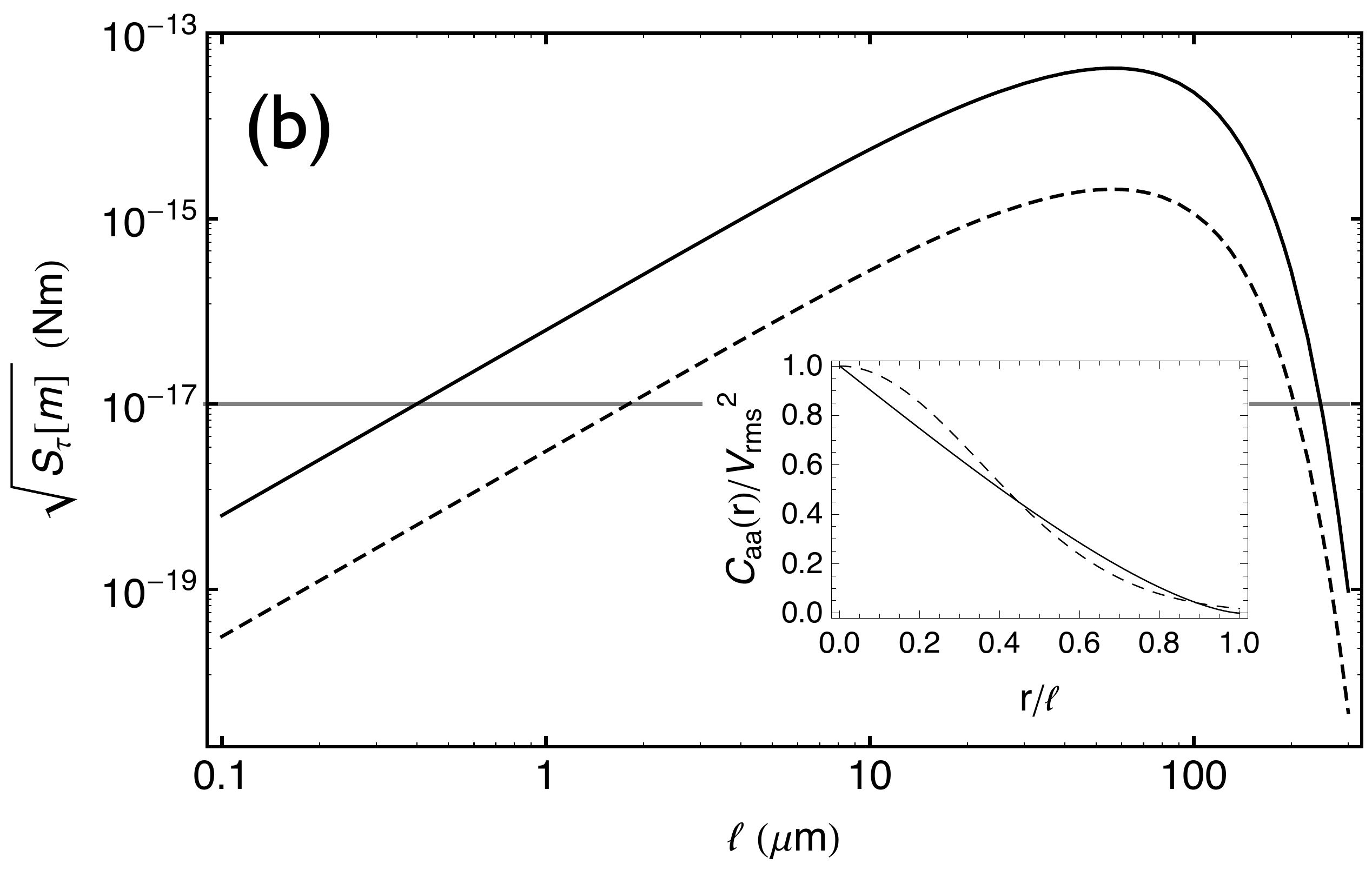} 
\caption{(Color online) (a) Set-up of the Birmingham experiment to measure non-Newtonian gravity with a torsion pendulum using the isoelectronic technique. 
For clarity, the gold coating on top of the mass-density modulations of both disks is not shown.  
The underside of the top disk is also density modulated.
(b) Root-mean-square amplitude of torque fluctuations as a function of typical patch size in the band relevant to gravity measurements in the experiment ($m=2048$ in Eq.(\ref{torque-power-spectrum})). The solid line is for $V_{\rm rms}=45$ mV (rms voltage for clean gold surfaces), and the dashed line is for $V_{\rm rms}=10$ mV (reduced rms voltage to mimic surface contamination).
The gray horizontal line represents the design sensitivity of the experiment.
The other parameters are $D=13 \; \mu$m  and $R_0=40$ mm. The inset shows the quasi-local voltage correlation function (solid) and the Gaussian correlation function (\ref{corr}) (dashed).}
\label{}
\end{center}
\end{figure}

The patch torque signal, although systematic, is random in the experiment. We quantify its variance by means of the patch torque power spectrum defined as 
\begin{align}
\label{}
 S_\tau[m] = \frac{1}{2\pi} \int_0^{2\pi} d \varphi \ e^{-i m \varphi} \langle \tau(\phi+\varphi) \tau(\phi) \rangle.
\end{align}
Using the above correlation function we obtain
\begin{eqnarray}
\label{torque-power-spectrum}
S_\tau[m] &=& \frac{1}{32} \pi^2 \varepsilon_0^2  m^2 R^2_0 V^4_{\rm rms} \ell^4  \int_0^\infty d k \frac{k^3 e^{ - \frac{1}{8} \ell^2 k^2}}{\sinh^2 kD}   \nonumber  \\
&& \times  [ J^2_m(k R_0) - J_{m+1}(k R_0) J_{m-1}(k R_0)] ,
\end{eqnarray}
where $J_m(x)$ is the $m$th order Bessel function of the first kind \cite{Gradshteyn} (see supplemental information for details). The relevant value of the integer $m$ is 2048 because this corresponds to the angular frequency of the mass-modulated signal.
Using this expression we compute the expected torque fluctuations as a function of patch size and compare with the design
sensitivity of  the Birmingham experiment (see Fig.1b).  
For clean gold surfaces the rms patch torque fluctuations exceed the design sensitivity over nearly the entire range of physical patch sizes, implying that patch effects dominate non-Newtonian signals. 
Surface contamination is known to lower the typical root-mean-square variations of surface patch potentials \cite{Rossi92}. 
We mimic this effect with a reduced $V_{\rm rms}$ in Fig. 1b, which shows that the patch signal can be lowered by a factor of 20 when the rms potential variation is decreased by a factor of 4.5. This suggests that purposefully contaminating the sample surfaces with adsorbates might beneficially suppress the patch torque power spectrum. 


{\it The IUPUI Experiment and Patch Force:}
In this experiment a Au-coated sapphire sphere (radius $R=150 \ \mu$m) is glued to a microtorsional oscillator (MTO) \cite{Ricardo-new-exp}. The sphere-MTO assembly is brought into close proximity to a rotating disk, as schematically shown in Fig.2a. The apex of the sphere is at a distance $L=1$ cm from the axis of rotation.
The disk is the source mass formed by $2 p$ ($p = 200$) sectors that subtend the same angle, which are alternatively made of Si and Au. All the sectors are covered by a uniform layer of Au (thickness 200 nm), thick enough to make the setup isoelectronic. The disk is set to rotate at an angular frequency $\Omega=\omega_r/p$, where 
$\omega_r \approx 2 \pi \times 300$ Hz is the resonant frequency of the MTO-sphere assembly. The experiment is designed to achieve a sensitivity of 0.1 fN with an integration time of 1000 s. The experimental procedure comprises measuring the normal force on the MTO at its resonant frequency $\omega_r$ as a function of the disk-sphere separation $D$.

\begin{figure}[t]
\begin{center}
\includegraphics[width=3in]{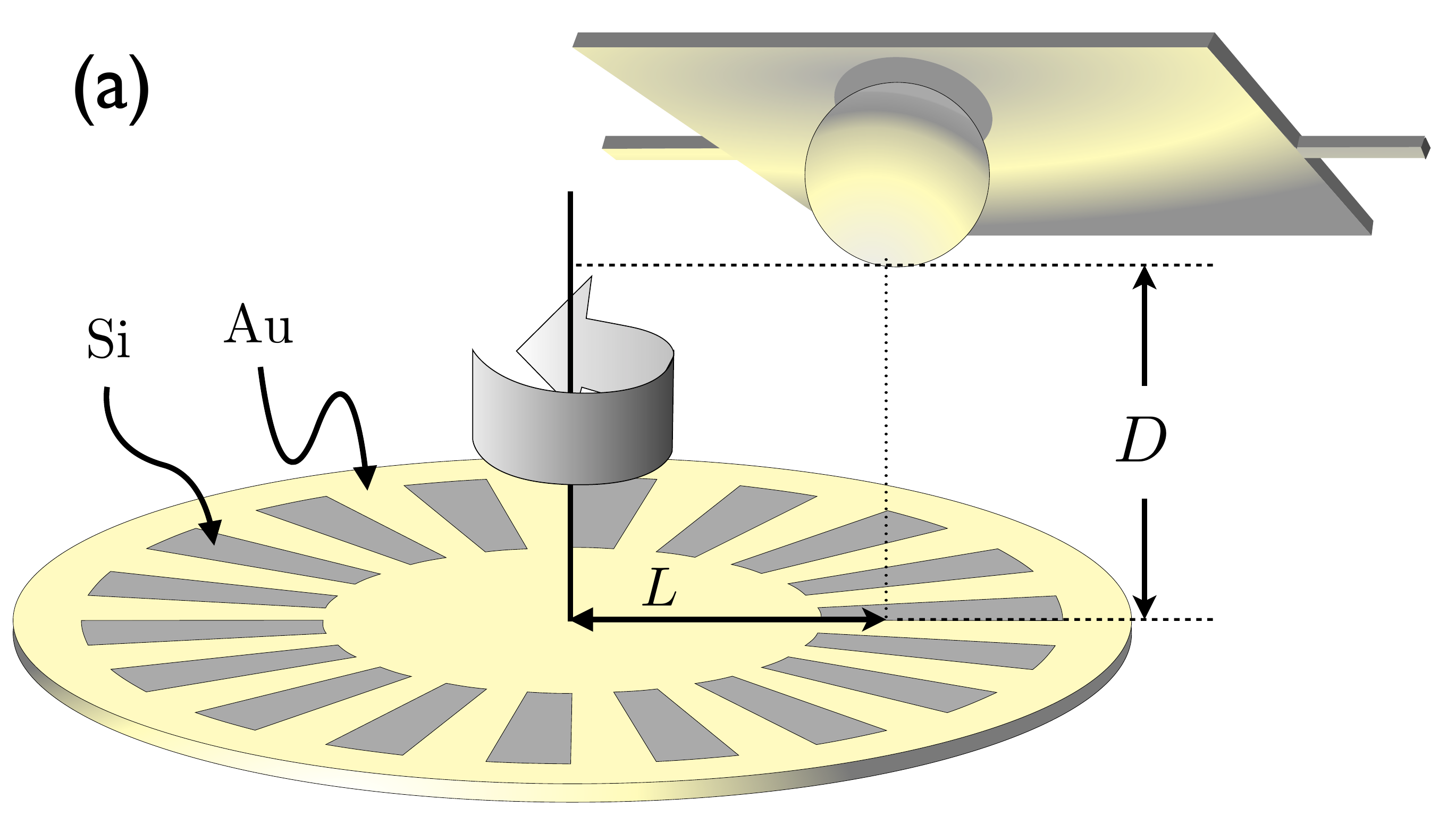}
\includegraphics[width=3.0in]{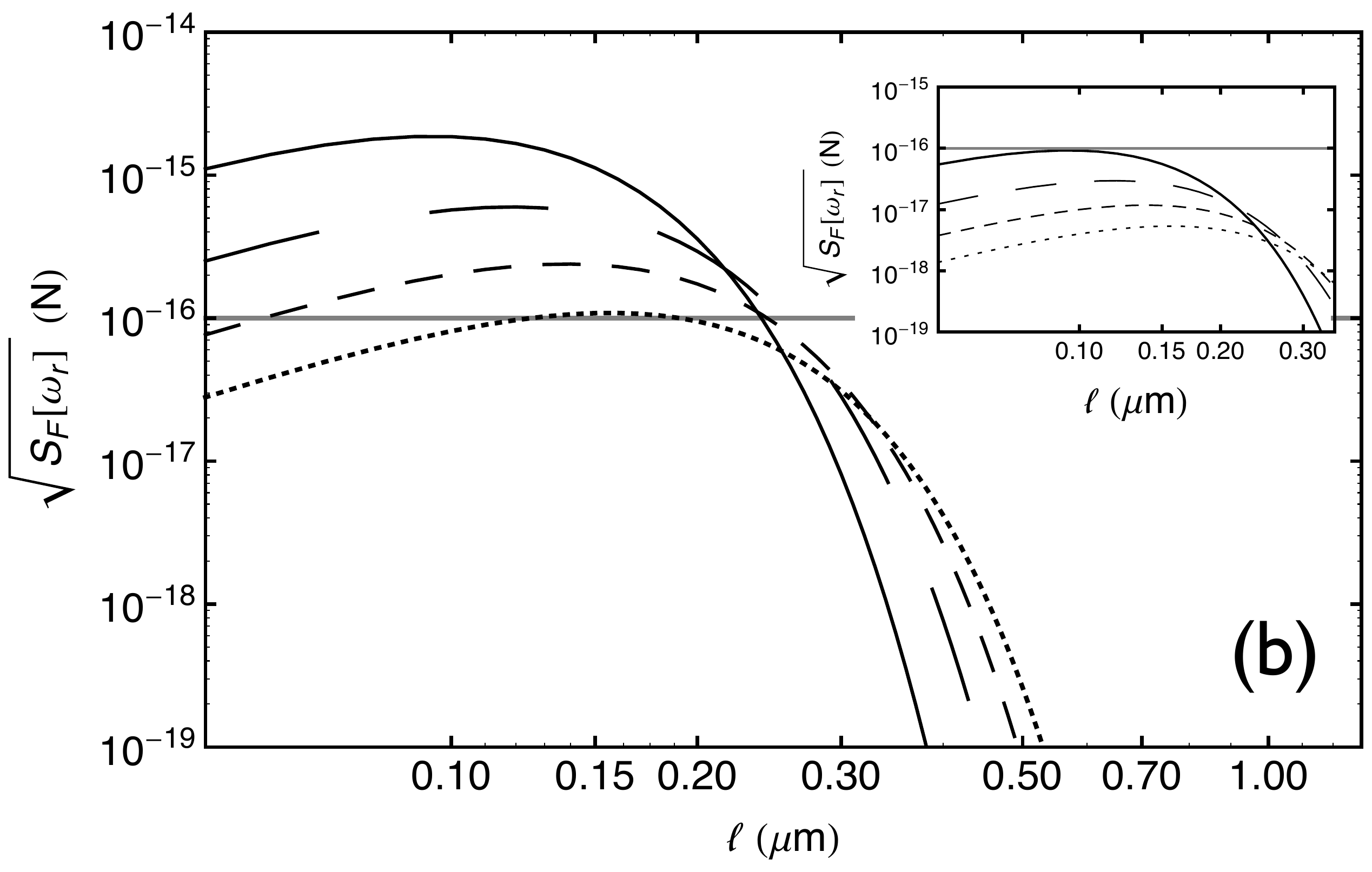}
\caption{(Color online) (a) Set-up of the IUPUI experiment to measure non-Newtonian forces with a torsional balance using the isoelectronic technique. 
For clarity, the gold coating on top of the mass density modulated disk is not shown.
(b) Expected force fluctuations in the band relevant to gravity signals in the IUPUI experiment  as a function of typical patch size for different separations $D$: $100$ nm (solid), $150$ nm (long-dashed), $200$ nm (short-dashed) and $250$ nm (dotted). $V_{\rm rms}$ is set  to $45$ mV. The gray horizontal line demarcates the force sensitivity in the IUPUI experiment at $0.1$ fN.
The inset approximates the effects of contamination which maintains the same convention as the main figure except $V_{\rm rms}$ is set to $10$ mV.}
\label{ }
\end{center}
\end{figure}

The Fourier component of the patch force at the resonance frequency of the MTO adds a systematic signal to the measurement.
We estimate the magnitude of this effect by considering the patch force signal as a function of the rotation angle and describe the surface potentials using
the approximate correlation function (\ref{corr}). For simplicity, we approximate the disk by an infinite plane, and treat the sphere's curvature using the proximity force approximation (PFA).
The latter assumes that the force between non-planar objects is the sum of the forces between infinitesimal planar sections, and relates the sphere-plane force $F$ to the plane-plane energy, $\mathcal{E}$, as $F = 2 \pi R \ \mathcal{E}/A$, where $A$ is the area of the plate.  
In \cite{Behunin12b} the validity of the PFA for describing patch effects in the sphere-plane geometry was tested and verified for $D \ll R$
and for a variety of patch sizes (see Fig. 4 in this reference). The plane-plane energy is given by

\begin{eqnarray}
\label{Energy-Ricardo}
\mathcal{E}(\phi)  =  \int_A d^2 x \int_A d^2 x'  V_1({\bf x}) 
 \Delta(\tilde{\bf x} - {\bf x}',D) V_2({\bf x}')  +  \ldots, \nonumber \\
\end{eqnarray}
where now $\tilde{\bf x}= R^{-1}_\phi \cdot ({\bf x} + L \hat{x}) - L \hat{x}$ and $\hat{x}$ is a unit vector pointing in the $x$-direction. The origin of the Cartesian coordinate system is centered on the point of closest separation between the sphere and the disk, so that the series of transformations leading to $\tilde{\bf x}$ represents a rotation of the plane on an axis centered at $-L \hat{x}$. The dots indicate terms independent of the rotation angle that do not contribute to the relevant patch signal. 
Note that the mean value of the angle-dependent energy is zero since $\langle V_1({\bf x}) V_2 ({\bf x}') \rangle = 0$.
 
The patch force power spectrum is defined as  
\begin{equation}
\label{power}
S_{F}[\omega] = \frac{1}{T}\int_{0}^{T} dt' \ e^{i \omega t'} \langle F(t+t') F(t) \rangle,
\end{equation}
where $T$ is the time period over which the measured signal is averaged, and
the time-dependence of the force arises from replacing $\phi$ with $\Omega t$
in (\ref{Energy-Ricardo}). 
The force power spectrum at the resonance frequency of the MTO is then given by
\begin{eqnarray}
&& S_{F}[\omega_r] =   \frac{\pi^3}{8}  \varepsilon_0^2 \frac{R^2 V^4_{\rm rms} \ell^4 }{A}     \int_0^\infty d k \frac{k^3 e^{ - \frac{1}{8} \ell^2 k^2}}{\sinh^2 kD}   \nonumber  \\
&& \times  [ J^2_{p}(k \sqrt{A/\pi}) - J_{p+1}(k \sqrt{A/\pi}) J_{p-1}(k \sqrt{A/\pi})] .
\label{force-power-spectrum}
\end{eqnarray}
Note that it depends upon the plate area, which is expected given that PFA can be viewed as an area
average of the plane-plane force (see supplementary information).  
The finite size of the sphere limits the sphere-plane interaction to an effective area  $A_{\rm eff}$ on the plane.
Within the PFA we can estimate it by equating the sphere-plane force to the product of the plane-plane pressure $P$
as $F = P A_{\rm eff}$. Using that ${\mathcal E}/A \sim D P$, it follows that $A_{\rm eff} = 2 \pi R D$. Therefore, the relevant
area in (\ref{force-power-spectrum}) is  $A=2\pi R D$. 
An important observation is that the patch force variance depends upon the rotation frequency (recall $p=\omega_r/\Omega$). Hence the 
influence of patches can be ruled out if one measures the variance of the force signal, assuming no other systematic signals
are present, for two different rotation frequencies and their ratio is unity. For example, the ratio of $S_F[\omega_r]$ for rotation frequencies $\omega_r/p$ and $\omega_r/(p+5)$ is $0.77$ at $D= 150$ nm and for $100$ nm patches. 
By taking the square root of $S_F[\omega_r]$ we can estimate the magnitude of the systematic patch force fluctuations at the frequency 
where the sought-after gravity signal should be present. In Fig.2b the magnitude of the patch signal is computed for several sphere-plate separations. We find that patch force signals are large enough to be measured for clean gold surfaces in the distance range $150-250$
nm and therefore may detrimentally limit constraints on hypothetical new forces.  The inset shows that contamination may prevent patches from influencing the IUPUI experiment altogether. 

A similar analysis can be performed for the previous isoelectronic experiment done at IUPUI in 2005 \cite{Decca05}. 
In this experiment the inhomogeneous source mass was deposited on a MTO and the sphere was moved from  
regions of high to low mass density and back, following a periodic motion $z(t) = D + \delta D \cos \omega_z t$ normal to the MTO
and $x(t) = x_0 \cos \omega_x t$ parallel to its axis. The modulation frequencies summed to the resonant frequency $\omega_r$ of the MTO. The experimental noise was  observed to be 0.3 fN with an integration time of 1000 s. The experiment measured a non-vanishing signal (see Fig. 3 of \cite{Decca05}), 
that was not attributed to a new hypothetical force but to a differential Casimir force arising from a small variation in the height of the coated
sample on the MTO. Even in the absence of such height variation, random (although systematic) patch forces could have contributed to the measured signal
in Fig. 3 of \cite{Decca05}. For this set-up  the patch force power spectrum at the MTO's resonance frequency is given by (see supplementary information)
\begin{eqnarray}
&& S_{F}[\omega_r] =   \frac{1}{4}  \varepsilon_0^2 \frac{R^2 \delta D^2}{A}     \int d^2 k \frac{k^4 \coth^2 kD}{\sinh^2 kD} J^2_1(k_x  x_0) C^2[{\bf k}],  \nonumber \\
\end{eqnarray}
where $C[{\bf k}]$ is the Fourier transform of (\ref{corr}) and, as before, $A=2 \pi R D$. In Fig. 3 we plot the magnitude of the expected patch force fluctuations as a function of patch size for different sphere-MTO distances. 
It is possible that the signal observed in \cite{Decca05} is due to patches.
The data of Fig.3 together with Fig. 3 of \cite{Decca05} can be used to constrain 
the possible patch sizes which could hypothetically produce 
the observed force. In order for our model to be compatible with 
the observations in \cite{Decca05} the patches must be smaller than  
$600$ nm for clean gold. 

\begin{figure}[t]
\begin{center}
\includegraphics[width=3.0in]{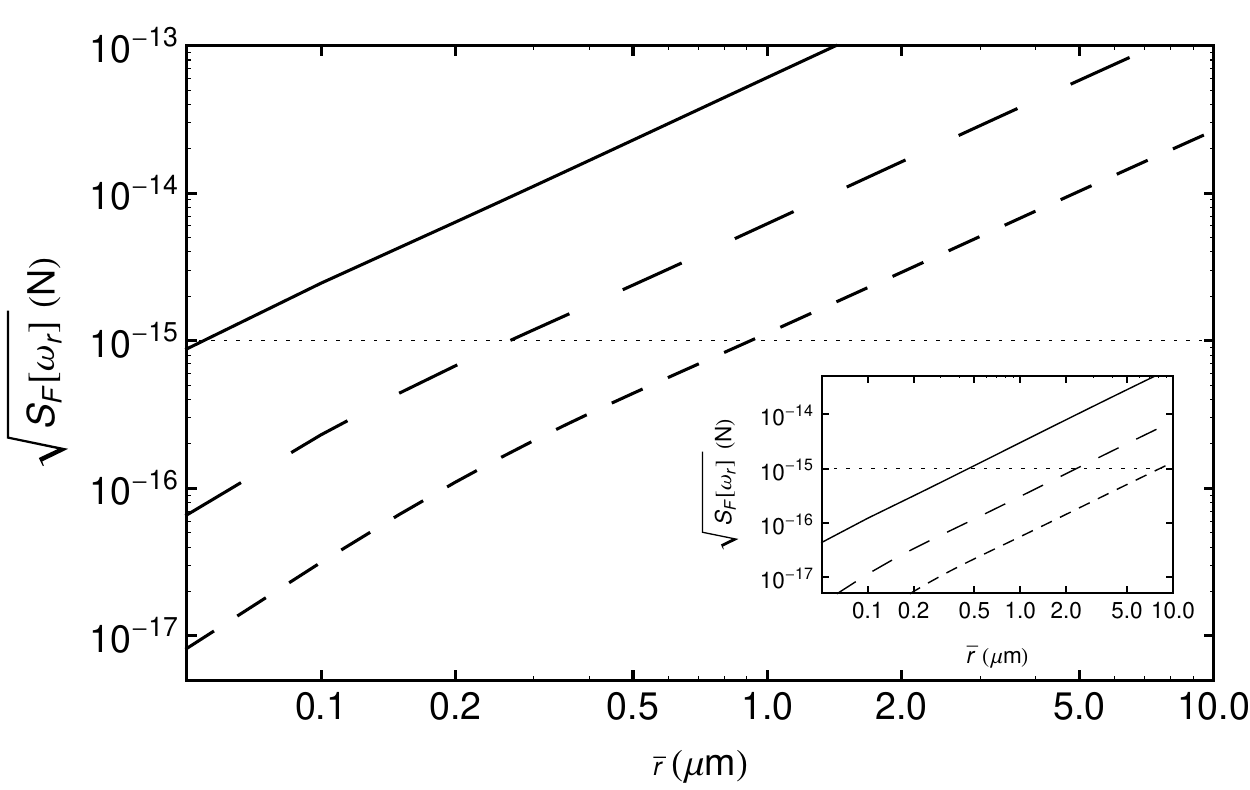}
\caption{Expected force fluctuations in the band relevant to gravity signals for the 2005 IUPUI experiment \cite{Decca05} as a function of typical patch size for different separations $D$:  from top to bottom $100$ nm, $250$ nm, and $500$ nm. 
The gray horizontal line indicates the force sensitivity for the experiment. The rms patch potential is set to that of clean gold, $V_{\rm rms}=45$ mV. The inset shows the expected patch force fluctuations at $150$ nm (dashed line) and
the black horizontal line represents the magnitude of the measured signal in Fig. 3 of \cite{Decca05} at this particular distance (as in the main
figure, the gray horizontal line is the force sensitivity). 
The constraints on forces reported in \cite{Decca05} imply that patches cannot have
sizes exceeding $600$ nm (corresponding to patch sizes right of the vertical line), otherwise they could result in patch forces larger than the total measured signal.
Other parameters are: $\delta D= 10$ nm, $x_0= 75 \ \mu$m, and $R= 50 \ \mu$m.
}
\end{center}
\end{figure}

In summary, we have developed the theory for patch effects relevant to isoelectronic experiments aimed at measuring
new hypothetical forces, in which the mean effects of patches are nullified by design, but whose fluctuations induce non-zero
systematic patch force and torque signals. We have found estimates for the root-mean-square amplitude of such fluctuations,
and quantified limits on the accuracy of two proposed isoelectronic set-ups. The application of our analysis to a previous isoelectronic experiment suggests the possibility that part of the force signal could have been due to the presence of patches on the samples. Our analysis indicates that the systematic effects of patches in all these experiments decrease for contaminated samples. 
Although the patch model employed in this work is inspired by physical arguments, the actual experimental patch distribution  need not be precisely described by this model. Consequently, our limits on accuracy to isoelectronic experiments are only qualitative, a quantitative analysis requires the measurement of the patch distribution for the given experimental samples.

RB and DARD acknowledge the support of LANL LDRD program. 
RSD acknowledges support from the IUPUI Nanoscale Imaging Center, Integrated Nanosystems Development Institute, and the Indiana University Center for Space Symmetries. CCS acknowledges support from STFC (UK).

\end{document}